# Automated source classification using a Kohonen network

P. H. Mähönen[1,2] and P. J. Hakala[1]

[1] *Department of Physics, Astrophysics, University of Oxford, Oxford, OX1 3RH, UK*[1]

[2] *Department of Theoretical Physics, University of Oulu, Linnanmaa, SF-90570, Finland*

**ABSTRACT**

We report progress in the development of automatic star/galaxy classifier for processing images generated by large galaxy surveys like APM. Our classification method is based on neural networks using the Kohonen Self-Organizing Map approach. Our method is novel, since it does not need supervised learning, i.e. human factor, in training. The analysis presented here concentrates on separating point sources (stars) from extended ones. Using simple numerical experiments we compare our method of image classification to the more traditional (PSF-fitting) approach of DAOFIND.

*Subject headings:* methods: data analysis – numerical – techniques: image processing

---

[1] e-mail: p.mahonen@physics.ox.ac.uk; p.hakala@physics.ox.ac.uk

## 1. Introduction

The present day deep surveys of stars and galaxies spread over a large area of the sky. The automated surveys like the APM catalogue (Automated Plate Measuring machine) and future surveys such as 2dF collaboration and the Sloan Digital Sky Survey will produce a vast amount of data to be analyzed. One of the major difficulties with these surveys is the discrimination between stellar and nonstellar objects, particularly faint magnitudes have been difficult to deal with using traditional discrimination methods. In this letter we introduce an *automated* algorithm for classification based on neural networks. Odewahn *et al.* (1992) have pioneered this approach using classical supervised learning networks, such as a backpropagation network, with the University of Minnesota Automated Plate Scanner for the Palomar Sky Survey. Our method is using unsupervised learning, which means that there is no human contribution at all. This ensures that the classification process is self-consistent, and if mistakes are made, they are always made in similar fashion. It should be also noted that the present method does not use any quantitative parametrization of the images (see e.g. Rhee 1990, Heydon-Dumbleton 1989). This makes the recognition process fast and reduces possible measuring errors.

## 2. Neural networks

Neural networks are information processing systems which share certain performance characteristics in common with biological neural networks (Lippmann 1987). In a typical network, information passes between neurons over connection links using weight sets which are unique to each neuron. These weights are then changed during the training. Typically an input vector is connected to *input layer*, subsequent layers are called a *hidden layers*, and output layer gives us results. Neural networks are known to be capable of performing complex pattern recognition tasks.

In this letter we are using the *Kohonen Self-Organizing Map* (Kohonen 1989,1990) scheme[2] known also as a topology-preserving map. Kohonen nets are single layer nets using unsupervised learning, where no *a priori* information of the input data is required. This feature of Kohonen's net is the fundamental difference between older and current neural network classifiers within astrophysics. Because the decision boundary of a self-organizing network is established without a training set consisting of input paterns and human-assigned class designations, there is no bias introduced by a human classifier at the time of training.

There are $M$ neurons in the net arranged in a one- or two-dimensional array. With every neuron, a parametric *weight vector* $\mathbf{w}_j$ is associated. The architecture of the Kohonen net is shown in Figure 1. In our case, the input data is a two-dimensional array (30×30 image) and we are using a 2-dimensional (20×20) network. Let $\mathbf{x} \in \mathcal{R}$ be an input data vector (image) and $\mathbf{w}_{ij}$ be a weight vector, connecting neuron $j$ to pixels $i$. During each self-organization step the net computes statistic $\mathbf{D}_j$ for each of the neurons.

$$\mathbf{D}_j = \sum_i ||\mathbf{x}_i - \mathbf{w}_{ij}||, \qquad (1)$$

where $||\mathbf{x}_i - \mathbf{w}_{ij}||$ is the Euclidean distance. Instead of Euclidean distance one can also use a dot product $\mathbf{x}_i \cdot \mathbf{w}_{ij}$. Next, the winning neuron is selected by choosing a neuron, whose $\mathbf{D}_j$ from (1) is largest(in case of dot product notation) or smallest (in case of Euclidean notation). We denote the index of this winning neuron by $c$. This is the neuron whose weights most closely resemble the given input image. For all neurons within a specified neighbourhood of neuron $c$, the weights are updated by a function

$$\mathbf{w}_{ij}(\text{new}) = \mathbf{w}_{ij}(\text{old}) + h_{ci}(\text{old})[\mathbf{x}_i - \mathbf{w}_{ij}], \quad (2)$$

---

[2] We refer to this as the Kohonen net for brevity.

where $h_{ci}$ is the *neighbourhood kernel*. In this work we use the Gaussian function

$$h_{ci} = \alpha(t) e^{-\frac{|r_c - r_i|^2}{2\sigma^2(t)}}, \quad (3)$$

where $\alpha$ is a scalar-valued "learning rate", $\sigma(t)$ defines the width of the kernel, $r_c$ and $r_i$ are the radius vectors of neurons $c$ and $i$, respectively and $t$ is the discrete time coordinate. Both $\alpha(t)$ and $\sigma(t)$ are slowly decreasing functions of time, and taken to be linear functions. Initialization of the network is done by assigning random values for the initial weights. It should be noted that the choice of $\sigma(t_0)$ and $\alpha(t_0)$ is not very critical, but affects the learning speed of the net. We have used $\sigma(t_0)$=2.0 and $\alpha(t_0)$=0.05, which are conventional values.

The actual training of the net consists of generating a number of images and presenting these to the net. After presenting each image the neuron activations are computed according to (1) and the weights of connections are updated according to (2). The learning rate and the kernel radius are then linearly lowered and the next image presented. This cycle is repeated until the net produces satisfactory classifications. In our tests around 2000 iterations were required. During the training one can estimate the "goodness" of the weight set by computing (and monitoring) the $\chi^2$ between the input image and the "winning weight set", $\mathbf{w}_c$, at each training step.

$$\chi^2 = \sum_i (\mathbf{x}_i - \mathbf{w}_{ic})^2, \quad (4)$$

In principle we can stop training when $\chi^2$-value ceases to decrease significantly. In practice this leads to a solution, where "code-images", or cluster mean images for each of the classes are (a) well localised and (b) smooth in appearance.

Once the net is trained, the weights of connections from neurons $j$ to pixels $i$ contain the cluster means for the classification. This means that weights for each neuron can be understood as an image associated with it. We refer to these "pseudo-images" as "code-images". One can think that code-image represents an average image mapped to this class. Three code images are plotted in figure 3 representing weight sets of three different neurons sensitive to point sources, extended sources with particular orientation and blank fields respectively. Each of the neurons in the trained net has its own code-image and together the code-images map the entire input space covered in the training set.

In our case training is a very simple process. We have $N \times N$ image, whose pixel values are mapped to smaller $M \times M$ net so that each pixel is connected to each of the neurons. Training is achieved by presenting each image in our training set to the Kohonen net, which is subsequently self-organizing by stochastic iteration.

## 3. Application of Kohonen nets to object classification

One of the most interesting features of Kohonen nets is the unsupervised learning scheme. The Kohonen net resembles statistical clustering algorithms (such as KMEANS) as it is capable of independently finding intrinsic clustering in the input parameter space. Thus no *a priori* information on the classification of training images is required.

As applied to source classification, the input parameter space of the Kohonen net consists of individual pixel values. For instance, an image of 30x30 pixels defines a 900 dimensional parameter space. Images with different characteristics (i.e. different types of objects) are presented to the net, which converges to yield the clustering hidden in the input images. For this study we used a net of 20x20 neurons, each of which was connected to each of the 900 input pixels (see also fig.1).

In order to train the net we need sample input images. In this study we have used simulated CCD-subframes (30x30 pixels) containing either a point source (represented by a Gaussian dis-tribution), an extended source (based on King model for elliptical galaxies) or just a sky background. In the case of extended sources the scale, orientation and inclination were also varied randomly. All of the images contained a Poisson (approximated by a Gaussian) background with a mean count rate of 100 counts/pixel. In addition to this, the source images contained either a point source or an extended source centered in the 30 × 30 frame. These where generated by taking random photons from an appropriate probability distribution associated with point sources (Gaussian) and extended sources (King model) and adding these to the sky background image. The images generated for the training typically contained 5000 counts (total) from the object in addition to the sky count rate of $\sim$ 90000 (30 × 30 pixels × $\sim$ 100 counts/pixel). These images were then mean subtracted and normalised to the unit length. One must bear in mind that although we are dealing with images, the net sees the input as a vector. This means that the real input to the net is a 900 element vector of unit length.

After training we tested the net with 2500 randomly generated input images similar to those used for training. We then examined the net responses to those images. The 'net response' means the neuron $c$ that produces the highest activation value $\mathbf{D}_c$ for a given input image. Figure 2. shows the net response (or $c$) distribution for 2500 test images containing point sources, extended sources and blank fields. One can easily notice that point sources systematically activate just few of the neurons, while different types of extended sources activate a 'ring of neurons' around the point sources. The rest of the neurons respond to blank fields. Furthermore, the extended sources are systematically classified within the 'ring', so that the fully trained net is able to estimate the orientation and extent of the extended sources. This aspect is more thoroughly investigated in a forthcoming paper (in preparation).

### 3.1. Point source detection

As an independent experiment we trained a Kohonen net with just simulated sky and point source images ignoring the extended sources. The idea was to compare the net performance in faint source detection against more traditional routines, such as DAOFIND.

We tested the trained net with a set of simulated images, 50 % of which contained a faint star in the middle (as in training) and the rest contained just a Poisson background. The object classification was determined by examining the net response areas sensitive for the two types of input images. These areas where predefined using the training set as an input for the fully trained net.

Each of the test images was presented to both the net and the IDL implementation of the DAOFIND routine. Again, each of these test images consisted of $\sim$ 100 counts/pixel background and a point source. The *total* number of counts from the point source was varied from 100 to 1000 counts. Figure 4. shows the correct classification probability of the Kohonen net (solid line) and the IDL DAOFIND (dotted line) as a function of point source count rate. It is easily seen that below $\sim$ 700 counts from the source the net clearly outperforms its rival.

### 4. Discussion

We have introduced the Kohonen net, which is capable of independently learning to discriminate between stellar and non-stellar objects. The tests are done using synthetic data as this is the only way to quantify the probability of the network finding a right class for an object. However, we tested the algorithm for true CCD data and in our forthcoming paper we use this network to scan a series of our NOT-observation CCD-frames and Palomar and APM survey images. This image scanning is done in two phases; first objects are found from images and then the neural net is used to classify each object. We will also report tests with problems such as poor image quality, blending etc. Another application we are presently developing is an automated galaxy morphology classifier using the Kohonen net approach. Lahav *et al.*(1995) and Storrie-Lombardi *et al.* (1992) have used neural nets for galaxy classification with very promising results. Hence we believe that the Kohonen net with its automatic classification may give some interesting insights to this problem. This method is superior at classifying faint objects and as an important added bonus, it is completely self-consistent in its' classification procedure. Since we are not using quantitative parameters of objects, such as elliplicity, the classification process is fast and robust. Measuring image parameters for neural nets would be more suitable for supervised learning schemes like backpropagation nets.

We would like to thank Gavin Dalton for useful discussions and suggestions, and invaluable help with manuscript. One of us (PM) acknowledges funding from the Academy of Finland (SERC).

---

This 2-column preprint was prepared with the AAS LaTeX macros v3.0.

Fig. 1.— The basic configuration of a Kohonen net. Each pixel is mapped to all neurons.

Fig. 2.— The distribution of maximum neural activation for (a) stellar, and (b) non-stellar objects, and (c) background images

Fig. 3.— The contour plot representation of three selected code-images. Panels (a,b,c) are same as in Fig. 2. These plots represent the weight sets of three neurons sensitive to the three different types of input images.

Fig. 4.— The solid and dotted lines show the correct classification probability for the Kohonen net and the IDL DAOFIND algorithm, respectively. One should note that with low luminosity objects the neural net is superior.

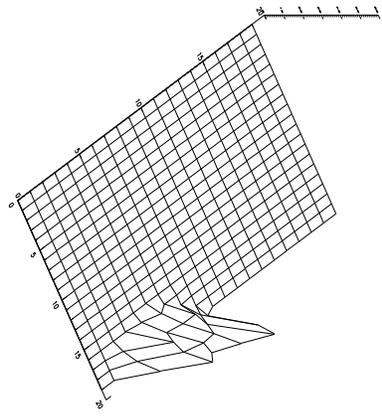
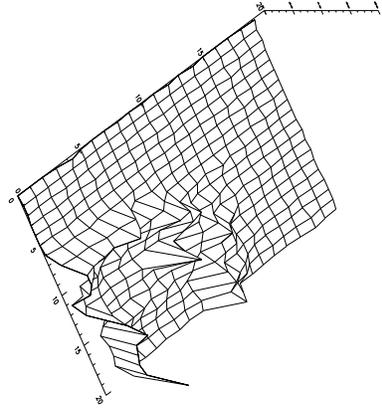
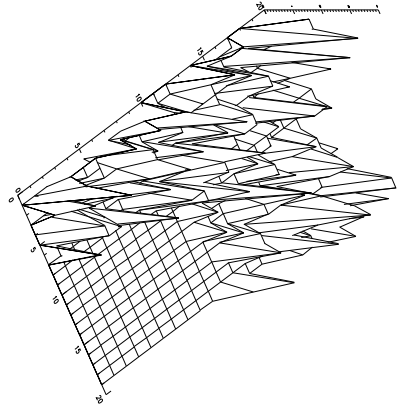

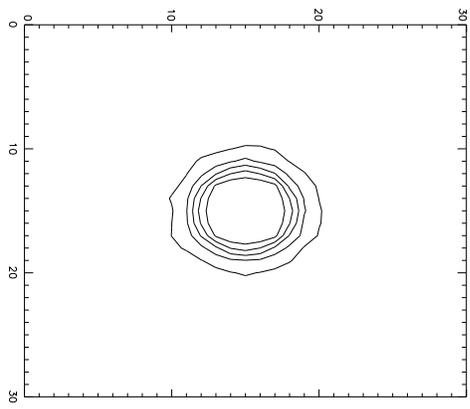

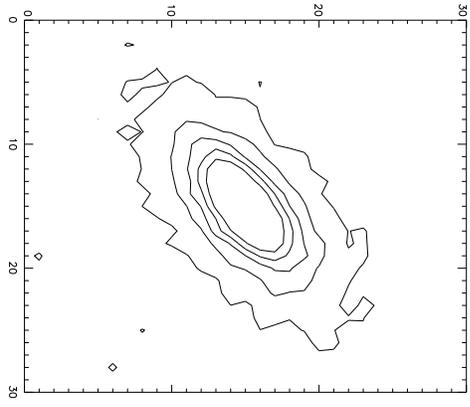

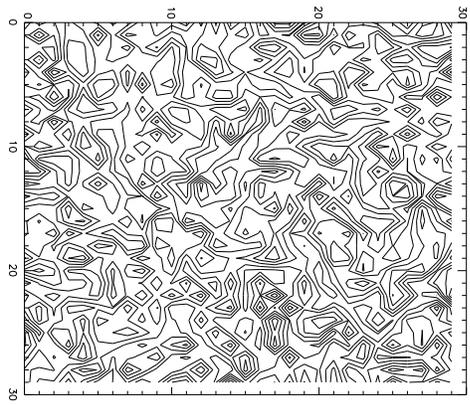

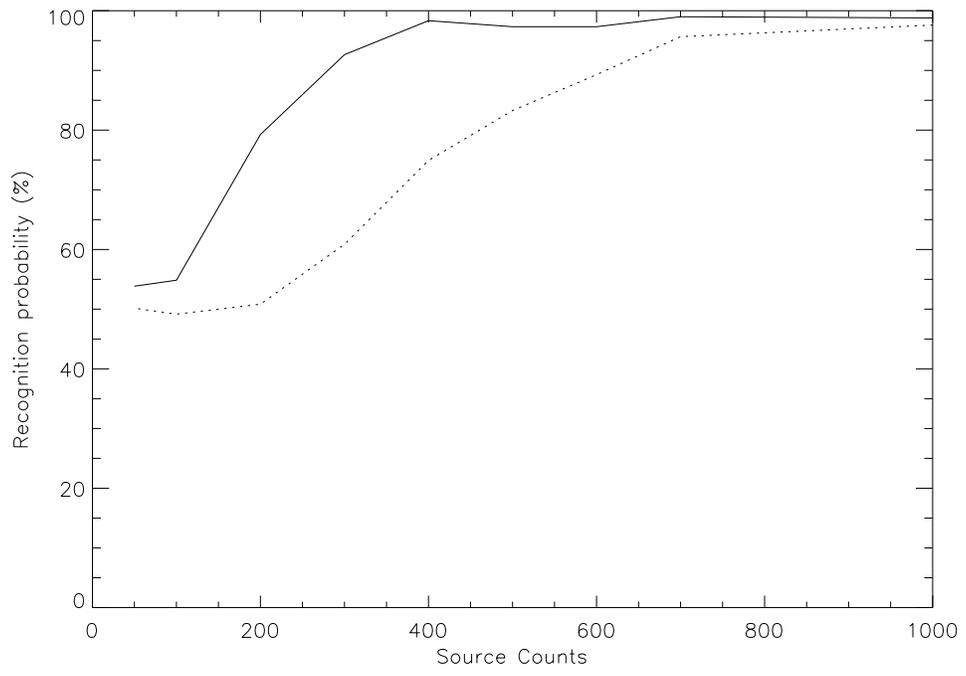